\documentstyle[aps,prc,psfig,epsfig]{revtex}            %PRC-Layout

\newcommand \be{\begin{eqnarray}}
\newcommand \ee{\end{eqnarray}}
                                               
\begin{document}
\draft
\twocolumn[\hsize\textwidth\columnwidth\hsize %Erzeugt zweiseitiges Layout im
           \csname @twocolumnfalse\endcsname  %richtigen PRC-Format (prc s.o.)

\title{Momentum conservation and local field corrections for the
  response of interacting Fermi gases}
\author{Klaus Morawetz$^{1}$, Uwe Fuhrmann$^{2}$}
\address{$^1$ LPC-ISMRA, Bld Marechal Juin, 14050 Caen
 and  GANIL, Bld Becquerel, 14076 Caen Cedex 5, France\\
$^2$ Fachbereich Physik, University Rostock, D-18055 Rostock,Germany}
\maketitle
\date{\today}
\maketitle
\begin{abstract}
We reanalyze the recently derived response function for interacting
systems in relaxation time approximation respecting density, momentum and energy conservation. We find
that momentum conservation leads exactly to the local field
corrections for both cases respecting only density conservation and
respecting density and energy conservation. This
rewriting simplifies the former formulae dramatically. We discuss the
small wave vector expansion and find that the response function shows
a high frequency dependence of $\omega^{-5}$ which allows to fulfill
higher order sum rules. The momentum conservation also resolves a
puzzle about the conductivity which should only be finite in
multicomponent systems.
\end{abstract}
\vskip2pc]

Recently the improvement of the response function in interacting quantum
systems has regained much interest \cite{RRWR99,MF99}. This quantity is important in a variety
of fields and describes the induced density variation if the system is
externally perturbed: $\delta n=\chi V^{\rm ext}$. 
As an example for an interacting system with potential $V$ the conductivity can be calculated from the response function via
\be
{\rm Re} \sigma=-{V\over 4 \pi} \omega {\rm Im} \chi.
\ee
One of the most fruitful concepts to improve the response functions
including correlations are the local field corrections $G$
\be
\chi={\chi_0\over 1+G \chi_0},
\ee
see
\cite{RRWR99,STLS68,M90} and references therein.

On the other hand there exists an extremely useful form of the response
function when the interactions are abbreviated in the relaxation time
approximation $\tau$ respecting density conservation\cite{Mer70}. One
of the advantages of the resulting Mermin formula (\ref{M-n}) is that it leads to the
Drude -like form of the dielectric function in the long wavelength limit
\be
\epsilon=1-V \chi=1- {\omega_p^2 \over \omega (\omega+{i\over \tau})}
\ee
with the plasma frequency $\omega_p$ for the Coulomb potential $V$ from which follows the conductivity
\be
{\rm Re} \sigma= {n e^2 \tau\over m (1+\omega^2 \tau^2)}=\left
  \{ \matrix{{n e^2 \tau\over m} +o(\omega)\cr
{n e^2\over m \omega^2 \tau}+ o({1\over \omega})} \right . .
\label{c1}
\ee
However one should
note that this formula is valid only for the extension to a
multicomponent system \cite{MWF97} (at
least a two-component system) since it makes no sense to speak of
conductivity in a single component system where the conductivity
should be infinite. Clearly the Mermin formula does not distinguish
these cases and cannot be sufficient to describe the
response. Therefore we will show that the inclusion of additional
momentum conservation will repair this defect (\ref{njwim}) and will lead to a
conductivity
\be
{\rm Re} \sigma= {n e^2 \tau\over m (1+\omega^2 \tau^2)}{n q^2\over m \omega^2}\left ({1\over \partial_\mu n}-{2 E\over n^2}\right
)
\ee
which shows indeed for the static limit a diverging behavior in
contrast to (\ref{c1}). 

There are two distinguishable cases, the single
component case where we have to include momentum conservation and
obtain divergent conductivity and the multicomponent case where we
should expect Mermin-like formulae in order to render the conductivity
finite. 
In order to bring these two extreme cases together the response
function for multicomponent systems should be considered \cite{MWF97}.

In this letter we want to restrict to the one - component situation.
In \cite{MF99} we have derived the density, current and energy response $\chi, \chi_J, \chi_E$ of an interacting quantum system 
\be
&&\left (\matrix {\delta n\cr\delta {\bf \nabla J}\cr \delta E} \right )=\left (\matrix{\chi \cr \chi_J \cr \chi_E}\right ) \,\,V^{\rm ext}\equiv{\cal X} \left (\matrix{1\cr 0\cr 0}\right ) V^{\rm ext}\equiv{\cal X}\nu^{\rm ext}
\nonumber\\
&&
\label{def}
\ee
to the external perturbation $V^{\rm ext}$ provided the density,
momentum and energy are conserved. The interacting system has
been described by the quantum kinetic equation for the density operator in relaxation time approximation 
where the relaxation is considered with respect to the local density
operator or the corresponding local equilibrium distribution function.
This local equilibrium is given by a local chemical potential $\mu$,
a local temperature $T$ and a local momentum $Q$ of mass motion. These
local quantities are specified by the requirement that the
expectation values for density, momentum and energy are the same when
calculated from local distribution function or performed with the density operator. 

The density response functions have been expressed in \cite{MF99} for the
inclusion of successively more conservation laws in terms of
polarization functions ${\cal P}=\{\Pi,\Pi_n,\Pi_E\}$ and have the general form
\be
{\cal X}={\cal P}(1-{\cal V} {\cal P})^{-1}
\label{gen}
\ee
due to the induced mean fields which can have density - and momentum -
dependent Skyrme form.  

When we note the free response function or Lindhard polarization
function without collisions as
\be
\Pi_0=s\int {d {\bf p}\over (2 \pi)^3} {f_0({\bf p}+{{\bf q}\over 2})-f_0({\bf p}-{{\bf q}\over 2}) \over {{\bf p q}\over m}-\omega -i0}
\label{pn}
\ee
with finite temperature Fermi functions $f_0$,
the inclusion of only density conservation leads to the Mermin
polarization \cite{Mer70}
\be
\Pi^{\rm n}({\bf q},\omega)&=&\frac{\displaystyle\Pi_0({\bf q},\omega+{i/\tau})}
 {\displaystyle1-\frac{1}{1-i\omega\tau}\left[1-\frac{\Pi_0({\bf q},\omega+
             {i/\tau})}{\Pi_0({\bf q},0)}\right]}\nonumber\\
&=&
(1-i \omega \tau){g_1(\omega+{i\over \tau}) g_1(0)\over h_1}           \label{M-n}.
\ee
 
If we include also the energy conservation we obtain \cite{MF99}
an additional term to (\ref{M-n})
\be
\Pi^{\rm n,E}(\omega)&=&(1-i \omega \tau)\left ({g_1(\omega+{i\over \tau}) g_1(0)\over h_1}
\right .\nonumber\\
&&\left . -\omega\tau i {(h_\epsilon g_1(0)-h_1 g_\epsilon(0))^2\over h_1(h_\epsilon^2-h_{\epsilon \epsilon} h_1)}\right )
\label{M-ne}
\ee
where we use the abbreviation
\be
h_\phi=g_\phi(\omega+{i\over \tau})-\omega \, \tau \, i \, g_\phi(0).
\label{h}
\ee

The different occurring correlation functions can be written in terms of moments of the usual Lindhard polarization function (\ref{pn})
as follows \cite{MF99}:
\be
g_1&=&\Pi_0,
\nonumber\\
g_\epsilon&=&- {n\over 2} +{m \omega^2 \over 2 q^2} \Pi_0 +{1\over 2 m}\tilde \Pi_2,
\nonumber\\
g_{\epsilon \epsilon}&=&-{7\over 6}  {E} -{n q^2 \over 16 m} (1+{4 m^2
  \omega^2 \over q^4})-{m^2 \omega^4 \over 4 q^4} \tilde \Pi_0
\nonumber\\
&&-{\omega^2 \over 2 q^2} \tilde \Pi_2-{1\over 4 m^2}\tilde\Pi_4.
\label{gg}
\ee
Integration via the chemical potential yields the higher moments of
the polarization function
\be
\tilde \Pi_2&=&2 m \int \limits_{-\infty}^\mu d\mu' \Pi_0,
\nonumber\\
\tilde \Pi_4&=&2 (2 m)^2 \int \limits_{-\infty}^\mu d\mu'\int \limits_{-\infty}^{\mu'} d\mu'' \Pi_0 
\label{ein}
\ee
and the density and energy are given by
\be
n&=&\int {d p\over (2 \pi \hbar )^3} f_0(p),\nonumber\\
E&=&\int {d p\over (2 \pi \hbar )^3} {p^2\over 2 m}f_0(p).
\ee

For the inclusion of additional momentum conservation to formulae
(\ref{M-n}) or (\ref{M-ne}) we obtain now a
tremendous simplification by observing that the formulae given in
\cite{MF99} can be rewritten as
\be
{1\over \Pi^{\rm n,j}(\omega)}-{1\over \Pi^{\rm n}(\omega)}&=&
{1\over \Pi^{\rm n,j,E}(\omega)}-{1\over \Pi^{\rm
      n,E}(\omega)}\nonumber\\
&=&- {i \omega \over \tau} {m\over n q^2}\equiv G.
\label{result}
\ee

This shows that the inclusion of momentum
conservation leads to nothing but the local field correction with the same form $G$ for both cases, the inclusion of only density
conservation {\it and} additional energy conservation. 
Formula (\ref{result}) is the main result of this paper since it leads
to a tremendous simplification. To see the advantages
more clearly we discuss now limiting cases.

The long wave length expansion is particularly important for the
classical limit and for the discussion of sum rules \cite{SM98}.
Since the discussion above has shown the advantage of discussing the
inverse polarization function instead of the polarization function
itself we proceed and give the expansion for the inverse polarization
functions (\ref{pn}), (\ref{M-n}) and (\ref{M-ne}) 
\be
{1 \over \Pi_0}&=&{m \omega^2 \over n q^2} -{2 E\over n^2}
+o(q^2),\nonumber\\
{1 \over \Pi^{\rm n}}&=&{m \omega (\omega +{i\over \tau}) \over n q^2}
-\left ({i\over \omega \tau} {1\over \partial_\mu n}+{2 E\over
    n^2}\right )
{\omega \over \omega +{i\over \tau}}+o(q^2),\nonumber\\
{1 \over \Pi^{\rm n,E}}&=&{1 \over \Pi_{\rm n}}
-{n q^2 \over 18 m}\left ({9\over \partial_\mu n}-{10 E\over n^2} \right )^2
{\omega \over (\omega +{i\over \tau})^3}+o(q^4).\nonumber\\
&&
\label{nn}
\ee
From equations (\ref{result}) it is straight forward to derive 
the expansions for $\Pi^{\rm nj}$ and $\Pi^{\rm njE}$.

The first observation is that up
to zeroth order in $q$ the local field corrections
(\ref{result}) induced by momentum conservation lead to an exact
cancellation of the effect of collisions in (\ref{M-n})  since we have
\be
{1 \over \Pi^{\rm n}}-{1 \over \Pi_0}=-G + o(q^0)
\ee
which shows that we have to go to the next order in $q$ as done in
(\ref{nn}).

Also one recognizes that the inclusion of energy conservation leads
only to corrections in next order of $q^2$ with respect to $\Pi^{\rm
  n}$. 
Moreover, we observe that this
correction even vanishes if we employ the zero temperature limit.
For zero temperature we have $E=3 \epsilon_f n/5$ and $\partial_\mu
n=3 n/2\epsilon_f$ with the Fermi energy $\epsilon_f$ such that 
\be
{1 \over \Pi^{\rm njE}}&=&{1 \over \Pi^{\rm nj}}+o(q^4)\nonumber\\
&=&{m \omega^2 \over n q^2} -{2 \epsilon_f\over
15 n} {9 \omega +{5 i\over \tau}\over \omega +{i\over \tau}}+o(q^4).
\ee 
Using (\ref{nn}) one can write all the effects of correlation
including conservation laws in one common local field factor
\be
\tilde G&=&{1 \over \Pi^{\rm n,j,E}}-{1 \over \Pi_{0}}\nonumber\\
&=&-{1\over
  1-i\omega \tau} \left ({1\over \partial_\mu n}-{2 E\over n^2}\right
)+o(q^2)\nonumber\\
&=&{1\over
  1-i\omega \tau} {8 \epsilon_f \over 15 n} +o(q^4)
\label{resul}
\ee
where the last line is valid for zero temperature.

This allows in turn to give the small wave vector expression of the
polarization function itself in  a Drude-like expression
\be
&&\lim_{{\bf q}\to 0}\Pi^{\rm n,j,E}({\bf q},\omega)=
\frac{n q^2}{m\omega\left[\omega+{n q^2\over m \omega}
{\rm Re}\tilde G(\omega)+{i \over \tilde\tau(\omega)}\right]}
                           \label{mmq1}\nonumber\\
&&
\ee
with the modified frequency--dependent
relaxation rate 
\be
\tilde\tau^{-1}={n q^2 \over m \omega}
{\rm Im}\tilde G(\omega)
\ee 
similar to \cite{RRWR99}. The advantage here is that we have simple explicit
formulae for the dynamical local field factor $\tilde G$ and the modified
relaxation rate while in \cite{RRWR99,RW98} this could only be given in
static approximation and involving complicated integrals.
If we had used simply the Mermin formula ({\ref{M-n}}) we would have
obtained $\tilde \tau =\tau$ and ${\rm Re}\tilde G=0$.

In particular we find for the
imaginary part
\be
\lim_{\omega\to\infty}{\rm Im}\Pi^{\rm n,j,E}({\bf q},\omega)&=&
-\frac{n^2 q^4}{\omega^5 \tau m^2}\left ({1\over \partial_\mu n}-{2 E\over n^2}\right
)\nonumber\\
&=&-\frac{8 \epsilon_f n q^4}{15\omega^5 \tau m^2},
\label{njwim}\\
\lim_{\omega\to\infty}{\rm Im}\Pi^{\rm n}({\bf q},\omega)&=&
-\frac{n^2 q^4}{\omega^3 \tau m^2}
\label{nwim} 
\ee 
showing a characteristic different high frequency behavior. While in
\cite{MF99} we have checked the improved convergence of first energy
weighted sum rule for the full expression (\ref{result}) we want to
point out that the $\omega^{-5}$ decrease for high frequencies allows
to fulfill higher order sum rules. The analytical discussion and
proof similar to \cite{SM98} will be devoted to a forthcoming work.

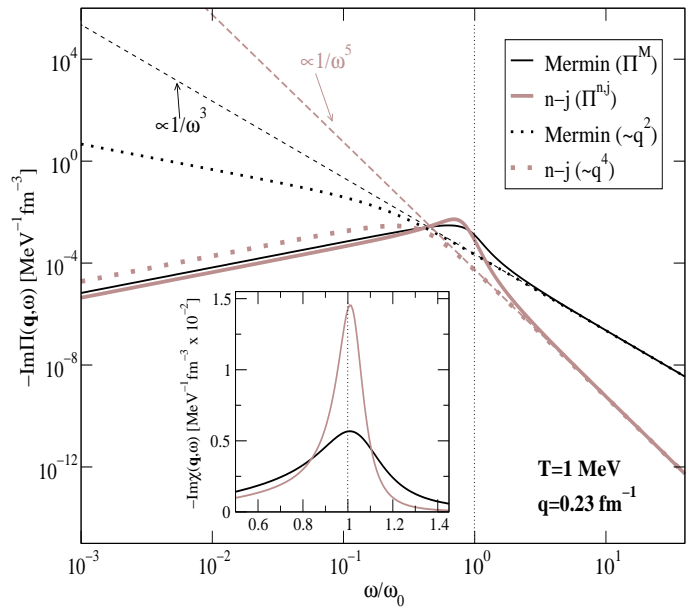
\begin{figure}[b]
\centerline{\psfig{file=compM_NP.eps,height=9cm,width=8cm,angle=-90}}
%/usr/users/roepke/uwe/Promotion/DISS/Teil3/compM_NP.eps,height=9cm,width=8cm,
%                                                        angle=-90}}
\caption{The imaginary part of the polarization function versus scaled
  energy for the
  Mermin formula (\protect\ref{M-n}) respecting only density
         conservation (black solid line) compared with the full
         expression (\protect\ref{result}) respecting
         energy, momentum and density conservation (gray solid line).
As an exploratory example 
         hot symmetric nuclear matter ($T=1$ MeV, $n_0=0.16{\rm
         fm}^{-3}$) with the
         wave vector $q=0.23 {\rm fm}^{-1}$ corresponding to $Pb$ has
         been chosen. Similar figures are obtained for plasma systems.
         The imaginary part of the response function is depicted in the
         inlay without logarithmic plot. The long wave length
         expansions for the Mermin (\protect\ref{M-n}) and the
         complete formula (\protect\ref{result}) are given by
         corresponding dotted lines. To guide the eye the high
         frequency limits (\protect\ref{nwim}) and (\protect\ref{njwim}) are given by long dashed lines.}
                      \label{compM_NP}
\end{figure}

In Fig. \ref{compM_NP}  we compare the imaginary part of the polarization 
function for the density and momentum approximation (gray lines) 
with the Mermin (density) approximation (dark lines) as function of energy
with the corresponding limiting cases. 

First we want to discuss the corresponding
complete expressions (solid lines) of the Mermin formula (\ref{M-n}) and the
formula (\ref{result}) including momentum, density and energy conservation. One recognizes that the low frequency limit agrees between
Mermin (density) formula and the complete formula while the high
frequency limit shows the characteristic different behavior of
$\omega^{-3}$ for Mermin (\ref{nwim}) and a stronger decrease of $\omega^{-5}$ for
the complete expression as have been seen in (\ref{njwim}). 
The high frequency expressions according to (\ref{njwim}) and (\ref{nwim}) are given by
corresponding dashed lines in the figure.

Let us now examine the long--wave length limits (\ref{mmq1}) 
of the Mermin formula
(\ref{M-n}) and the one including momentum, density
and energy conservation (\ref{result}) plotted in the figure as 
dotted lines. We see that the long wave limit of the Mermin formula
approximates the high frequency behavior of the Mermin formula (\ref{M-n}) nicely but fails for low
frequencies. In contrast, the long wave length expansion of the
expression including momentum conservations (\ref{mmq1}) shows an
excellent agreement with the complete expression (\ref{result}) for
both the high and low frequency limit. 
Please
remember that in the latter expression (\ref{mmq1}) the corrections of
order $q^2$
drop out and it is effectively of the order $q^4$. The nice numerical
agreement of the expression (\ref{mmq1}) with the full result (\ref{result})
underlines also the force of
local field corrections in constructing approximate formulae for the
response functions.

%--------------------------Fig. 1 --------------------------------------------

The valuable comments by J. D. Frankland
(GANIL) are gratefully acknowledged.

%\bibliography{kmsr,kmsr1,kmsr2,kmsr3,kmsr4,kmsr5,kmsr6,kmsr7,delay2,spin,refer,gdr,delay3}
%\bibliographystyle{prsty}

\end{document}